\documentclass[aps,prl,twocolumn,amsmath,amssymb]{revtex4-2}

\usepackage{float}
\usepackage{mathrsfs}
\usepackage{amsmath}
\usepackage{bm}
\usepackage{comment}
\usepackage{graphicx}
\usepackage{color}
\usepackage{overpic}
\usepackage{hyperref}
\usepackage{makecell}
\usepackage{etoolbox,xspace}

\newcommand{\jpsi}{J/\psi}
\newcommand{\Lambdabar}{\bar{\Lambda}}
\newcommand{\pbar}{\bar{p}}
\newcommand{\nbar}{\bar{n}}
\newcommand{\pip}{\pi^{+}}
\newcommand{\pim}{\pi^{-}}
\newcommand{\piz}{\pi^{0}}

\newcommand{\az}{\alpha_{0}}
\newcommand{\azbar}{\bar{\alpha}_{0}}
\newcommand{\am}{\alpha_{-}}
\newcommand{\alp}{\alpha_{+}}
\newcommand{\BESIIIorcid}[1]{\href{https://orcid.org/#1}{\hspace*{0.1em}\raisebox{-0.45ex}{\includegraphics[width=1em]{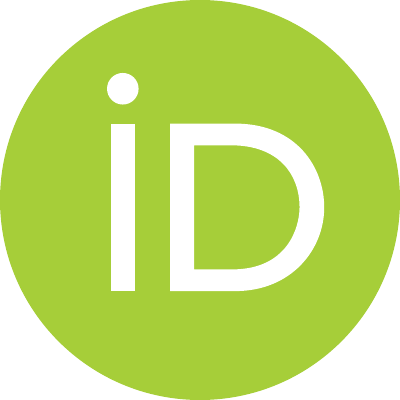}}}}

\begin{document}

\title{Test of $CP$ Symmetry in the Neutral Decays of $\Lambda$ via $J/\psi\to\Lambda\Lambdabar$}

%% Saved at => 2025-07-18
\author{
%% Saved at => 2025-07-18
M.~Ablikim$^{1}$\BESIIIorcid{0000-0002-3935-619X},
M.~N.~Achasov$^{4,c}$\BESIIIorcid{0000-0002-9400-8622},
P.~Adlarson$^{80}$\BESIIIorcid{0000-0001-6280-3851},
X.~C.~Ai$^{85}$\BESIIIorcid{0000-0003-3856-2415},
R.~Aliberti$^{37}$\BESIIIorcid{0000-0003-3500-4012},
A.~Amoroso$^{79A,79C}$\BESIIIorcid{0000-0002-3095-8610},
Q.~An$^{76,62,\dagger}$,
Y.~Bai$^{60}$\BESIIIorcid{0000-0001-6593-5665},
O.~Bakina$^{38}$\BESIIIorcid{0009-0005-0719-7461},
Y.~Ban$^{48,h}$\BESIIIorcid{0000-0002-1912-0374}
% ...
\\
\vspace{0.2cm}
(BESIII Collaboration)\\
\vspace{0.2cm} {\it
$^{1}$ Institute of High Energy Physics, Beijing 100049, People's Republic of China\\
% ...
$^{4}$ Budker Institute of Nuclear Physics SB RAS (BINP), Novosibirsk 630090, Russia\\
% ...
$^{37}$ Johannes Gutenberg University of Mainz, Johann-Joachim-Becher-Weg 45, D-55099 Mainz, Germany\\
$^{38}$ Joint Institute for Nuclear Research, 141980 Dubna, Moscow region, Russia\\
% ...
$^{48}$ Peking University, Beijing 100871, People's Republic of China\\
% ...
$^{60}$ Southeast University, Nanjing 211100, People's Republic of China\\
% ...
$^{62}$ State Key Laboratory of Particle Detection and Electronics, Beijing 100049, Hefei 230026, People's Republic of China\\
% ...
$^{76}$ University of Science and Technology of China, Hefei 230026, People's Republic of China\\
% ...
$^{79}$ University of Turin and INFN, (A)University of Turin, I-10125, Turin, Italy; (B)University of Eastern Piedmont, I-15121, Alessandria, Italy; (C)INFN, I-10125, Turin, Italy\\
$^{80}$ Uppsala University, Box 516, SE-75120 Uppsala, Sweden\\
% ...
$^{85}$ Zhengzhou University, Zhengzhou 450001, People's Republic of China\\

\vspace{0.2cm}
$^{\dagger}$ Deceased\\
% ...
$^{c}$ Also at the Novosibirsk State University, Novosibirsk, 630090, Russia\\
% ...
$^{h}$ Also at State Key Laboratory of Nuclear Physics and Technology, Peking University, Beijing 100871, People's Republic of China\\
% ...
}
%% ends here %%

}

\date{\today}

\begin{abstract}
Using $(10087\pm44)\times10^{6}$ $J/\psi$ events collected with the BESIII detector, 
a full angular distribution analysis is carried out on the process  $\jpsi\rightarrow\Lambda\Lambdabar\rightarrow n\piz\pbar\pip+c.c.$ 
The decay parameters $\alpha_{0}$ for $\Lambda\rightarrow n\piz$ and $\bar{\alpha}_{0}$ for $\Lambdabar\rightarrow \nbar\piz$ are measured to be $0.668\pm0.007\pm0.002$ and $-0.677\pm0.007\pm0.003$, respectively, 
yielding the most precise test for $CP$ symmetry of neutral decays of $\Lambda$, $A_{CP}^{0}=(\alpha_{0}+\bar{\alpha}_{0})/(\alpha_{0}-\bar{\alpha}_{0})$, to be $-0.006\pm0.007\pm0.002$.
The ratios $\alpha_{0}/\alpha_{-}$ and $\bar{\alpha}_{0}/\alpha_{+}$ are determined to be $0.884\pm0.013\pm0.006$ and $0.885\pm0.013\pm0.004$, 
where $\alpha_{-}$ and $\alpha_{+}$ are the decay parameters of $\Lambda\rightarrow p\pim$ and $\Lambdabar\rightarrow\pbar\pip$, respectively.
The ratios, found to be smaller than unity by more than $5\sigma$, confirm the presence of the $\Delta I = 3/2$ transition in the $\Lambda$ and $\Lambdabar$ decays, 
which is expected to improve the theoretical calculations for strong and weak phases, and $A_{CP}$, in hyperon decays.
In all results, the first and second uncertainties are statistical and systematic, respectively.
\end{abstract}

\maketitle

%% Introduction
The violation of the combined charge-conjugation ($C$) and parity ($P$) symmetries, known as $CP$ violation, 
is a key ingredient in explaining the observed matter–antimatter asymmetry of the universe through baryogenesis~\cite{Sakharov1991}. 
Despite accommodating $CP$ violation via the Kobayashi-Maskawa phase~\cite{Cabibbo1963, Kobayashi1973},the Standard Model (SM) fails to explain the baryon asymmetry of the universe~\cite{Barr1979}. 
Its predicted $CP$ violation is at least ten orders of magnitude too small.
%Although the Standard Model (SM) accommodates $CP$ violation with the Kobayashi-Maskawa phase~\cite{Cabibbo1963, Kobayashi1973}, 
%however, the magnitude predicted is at least ten orders of magnitude smaller than what is required to account for the observed baryon asymmetry~\cite{Barr1979}.
Experimental evidence for $CP$ violation has been found in $K$~\cite{CPVinKaon}, $B$~\cite{CPVinBofBABAR, CPVinBofBelle} and $D$~\cite{CPVinD} meson systems 
and the $\Lambda_{b}^{0}$ baryon system~\cite{CPVinLamb}, all these measurements are consistent with the SM predictions.
%Experimentally, $CP$ violation has been observed in $K$~\cite{CPVinKaon}, $B$~\cite{CPVinBofBABAR, CPVinBofBelle} and $D$~\cite{CPVinD} meson systems as well as in the $\Lambda_{b}^{0}$~\cite{CPVinLamb} baryon system, all of which are consistent with SM predictions. 
Therefore, to resolve the baryogenesis puzzle, additional sources of $CP$ violation beyond the SM must be considered. 
Hyperon decays, in particular, provide a sensitive probe for such effects~\cite{Salone2022, Donoghue1986}.

In the nonleptonic decays of spin $1/2$ hyperons, $CP$ violation arises from %the difference of amplitudes between the hyperon and antihyperon in their parity-violating two-body weak decays~\cite{Lee1957}.
the interference between parity-conserving ($P$-wave) and parity-violating ($S$-wave) amplitudes with differing $CP$-odd weak phases~\cite{Lee1957}.
In these decays, the angular distribution of the daughter baryon is given by $1+\alpha_{Y}\bm{P_{Y}}\cdot\bm{\hat{p}_{d}}$, where $\alpha_{Y}$ is the hyperon decay parameter, 
$\bm{P_{Y}}$ is the hyperon polarization vector, and $\bm{\hat{p}_{d}}$ is the unit vector along the daughter baryon momentum in the hyperon rest frame.
A $CP$-odd asymmetry parameter is defined as $A_{CP}=(\alpha_{Y}+\alpha_{\bar{Y}}) / (\alpha_{Y}-\alpha_{\bar{Y}})$, 
where $\alpha_{Y}$ and $\alpha_{\bar{Y}}$ denote the decay parameters of the hyperon and anti-hyperon, respectively. A nonzero $A_{CP}$ indicates $CP$ violation.
%The BESIII experiment has achieved the most precise measurement of hyperon CP violation to date, with a precision of $5\times 10^{-3}$, by analyzing entangled $\Lambda$ - $\Lambdabar$ pairs~\cite{Ablikim2022b}.
%However, the prediction of the SM reaches $-0.5\times 10^{-4}$, which is two orders of magnitude smaller than the experimental precision\cite{Donoghue1986}.

Within the SM, phenomenological studies based on different scenarios~\cite{Donoghue1986, Tandean2003, Salone2022} predict an extremely small $CP$ asymmetry 
in the $\Lambda$ nonleptonic two-body weak decay, $|A_{CP}^{\Lambda, \rm SM}|\leq 3\times10^{-5}$.
In contrast, several beyond SM theories predict significantly larger values, e.g. $|A_{CP}^{\Lambda, \rm MSM}|\leq 10^{-4}$ with minimal extensions of the SM (MSM)~\cite{PhysRevD.52.5257},
$|A_{CP}^{\Lambda, \rm CMPC}|\leq 7\times10^{-4}$ with chromomagnetic-penguin contributions (CMPC)~\cite{HE20221840},
$|A_{CP}^{\Lambda, \rm SUSY}|\leq 10^{-3}$ with supersymmetric models (SUSY)~\cite{PhysRevD.61.071701}, and 
$|A_{CP}^{\Lambda, \rm DM}|\leq 1.6\times10^{-3}$ with dark matter models (DM)~\cite{he2025btoksfinvisibledarkmatter}.
The main limitation in the precision of the theoretical predictions arises from the well known $S/P$-wave puzzle in hyperon nonleptonic decays,
where the $S$- and $P$-wave amplitudes cannot be simultaneously described within leading-order chiral perturbation theory~\cite{Jenkins:1991bt, Shi:2022dhw}.
Improved precision in the decay parameters can help resolve this puzzle.
%A dedicated experimental hunting for $CP$ violation in this decay is therefore desirable.
%Improvement in the precision of the decay asymmetry parameters can help in efforts to resolve the well known $S/P$-wave puzzle in hyperon nonleptonic decays,
%where the $S$- and $P$-wave amplitudes cannot be simultaneously well-described in leading-order chiral perturbation theory~\cite{Jenkins:1991bt, Shi:2022dhw}.
Such measurements can also sharpen the status of the $\Delta I = 1/2$ rule in the spin 1/2 hyperon sector~\cite{PhysRev.184.1663, PhysRevLett.24.843},
in comparison to kaon decays $K\to\pi\pi$~\cite{PhysRevLett.83.22, 1999335} 
where the $\Delta I = 3/2$ transitions are strongly suppressed relative to the $\Delta I = 1/2$ ones~\cite{PhysRevD.107.052008, Manzari:2020eum}.
With increasing experimental precision, $\Delta I = 3/2$ contributions can no longer be neglected~\cite{Ablikim2024} and must be included in theoretical calculations of $A_{CP}$.

Recently, BESIII has performed a series of precision studies of nonleptonic two-body hyperon decays~\cite{BESIII:2018cnd, Ablikim2022b, BESNature, Ablikim2024, PhysRevD.108.L031106, PhysRevLett.131.191802, PhysRevLett.133.101902, BESIII:2023ldd}.
The decay parameters of the charged $\Lambda$ decay mode $\Lambda\to p\pim$  ($\alpha_{-}$) and $\Lambdabar\to \pbar\pip$  ($\alpha_{+}$) 
were measured with the highest precision in the decay sequence $\jpsi\to \Lambda(\to p \pim) \Lambdabar (\to\pbar\pip)$~\cite{Ablikim2022b}.
%The decay parameters of $\Lambda$ charged decay mode $\Lambda\to p\pim$  ($\alpha_{-}$) and that of $\Lambdabar\to \pbar\pip$  ($\alpha_{+}$) are precisely measured in both subsequent decays of $\jpsi\to \Lambda(\to p \pim) \Lambdabar (\to\pbar\pip)$~\cite{Ablikim2022b},
%$\jpsi\to\Xi^{0}\bar{\Xi}^{0}\to\Lambda(\to p \pim)\piz\Lambdabar(\to \pbar \pip)\piz$~\cite{PhysRevD.108.L031106} and
%$\jpsi\to \Xi^{-} \bar{\Xi}^{+}\to \Lambda (\to n \piz) \pim \Lambdabar (\to \pbar \pip) \pip + c.c.$~\cite{Ablikim2024},
%individually, and are of the highest precision among all those of hyperons non-leptonic decays.
For the neutral decay modes of $\Lambda$, $\alpha_0$ for $\Lambda\to n\piz$ and $\bar{\alpha}_0$ for $\Lambdabar\to \nbar \piz$ 
were measured in the decay sequence $\jpsi\to \Xi^{-} \bar{\Xi}^{+}\to \Lambda (\to n \piz) \pim \Lambdabar (\to \pbar \pip) \pip$ and its charge conjugated channel~\cite{Ablikim2024}.
$\bar{\alpha}_0$ was also measured in $\jpsi\to \Lambda(\to p \pim) \Lambdabar (\to\nbar\piz)$ based on 1.3 billion $\jpsi$ events~\cite{BESIII:2018cnd}, 
but its precision is poor due to limited statistics.
%The decay parameters of $\Lambda$ charged decay mode $\Lambda\to p\pim$  ($\alpha_{-}$) and that of $\Lambdabar\to \pbar\pip$  ($\alpha_{+}$) are precisely measured  in both subsequent decays of $\jpsi\to \Lambda(\to p \pim) \Lambdabar (\to\pbar\pip)$~\cite{BESIII:2018cnd, Ablikim2022b},
%$\jpsi\to\Xi^{0}\bar{\Xi}^{0}\to\Lambda(\to p \pim)\piz\Lambdabar(\to \pbar \pip)\piz$~\cite{PhysRevD.108.L031106},
%$\jpsi\to \Xi^{-} \bar{\Xi}^{+}\to \Lambda (\to n \piz) \pim \Lambdabar (\to \pbar \pip) \pip + c.c.$~\cite{Ablikim2024} with 10 billion $\jpsi$ events
%and $\jpsi\to \Xi^{-} \bar{\Xi}^{+}\to \Lambda (\to p \pim) \pim \Lambdabar (\to \pbar \pip) \pip$ with 1.3 billion $\jpsi$ events~\cite{BESNature}, 
%individually, and are of the highest precision among all those of hyperons non-leptonic decays.
%However, those of $\Lambda$ neutral decay modes, $\alpha_0$ for $\Lambda\to n\piz$ and $\bar{\alpha}_0$ for $\Lambdabar\to \nbar \piz$, are only measured in the subsequent decays $\jpsi\to \Xi^{-} \bar{\Xi}^{+}\to \Lambda (\to n \piz) \pim \Lambdabar (\to \pbar \pip) \pip + c.c.$~\cite{Ablikim2024} so far.
%and $\jpsi\to \Lambda(\to p \pim) \Lambdabar (\to\nbar\piz)$ with 1.3 billion $\jpsi$ events~\cite{BESIII:2018cnd} so far. 
However, the uncertainties in $\alpha_0$ and $\bar{\alpha}_0$ remain substantially larger than those for the charged modes, motivating further improvement.
The decay $\jpsi\to\Lambda(\to n\piz)\Lambdabar(\to \pbar\pip)$ and its charge conjugated channel offer larger signal yields compared to the previous mode $\jpsi\rightarrow\Xi^{-}\bar{\Xi}^{+}$. 
Although the reduced quantum entanglement in this channel partly offsets the statistical advantage, it still provides a pathway toward more precise measurements of $\alpha_{0}$ and $\bar{\alpha}_{0}$. 
%and $\jpsi\to \Lambda(\to p \pim) \Lambdabar (\to\nbar\piz)$ with 1.3 billion $\jpsi$ events~\cite{BESIII:2018cnd} so far. 
%However, the detection and Monte Carlo (MC) simulation of the neutron and anti-neutron are a challenge for the accurate measurement that required detailed studies.

In this Letter, we present a full angular analysis of the decay $\jpsi\rightarrow\Lambda(\to n\piz) \Lambdabar (\to \pbar\pip)$ and its charge conjugate, 
based on $(10087\pm44)\times10^{6}$ $\jpsi$ events~\cite{Ablikim2022} collected with the BESIII detector.
The transversely polarization of $\Lambda$ and $\Lambdabar$ together with their spin correlation,
allow the angular distributions of both production and decay processes to be described by an extended helicity formalism~\cite{Perotti2019}.
%various properties are determined by an extended formalism that completely describes the angular distributions of the production and decay processes~\cite{Perotti2019}.
Unless otherwise indicated, the $\Lambdabar\to\nbar\piz$ mode is implicitly included throughout this text.

The BESIII experiment provides a unique platform to study hyperon production and decay properties in $e^{+}e^{-}$ annihilation to a $\Lambda\Lambdabar$ pair via the intermediate $\jpsi$ resonance.
%where the above challenges can be well addressed.
%The decay parameters and their corresponding $A_{CP}$ can be measured precisely due to the $\Lambda$ and $\Lambdabar$ being in a quantum-entangled state with  correlated spin and transversely polarized.
The decay chain $\jpsi\rightarrow\Lambda (\to n\piz)\Lambdabar (\to \pbar\pip) $ 
is characterized by five helicity angles $\xi=(\theta_{\Lambda}, \theta_{n}, \phi_{n}, \theta_{\pbar}, \phi_{\pbar})$ 
and the four parameters $\omega=(\alpha_{\jpsi}, \Delta \Phi, \alpha_{0}, \alpha_{+})$,
where $\theta_{\Lambda}$ is the polar angle of $\Lambda$ in the center-of-mass (c.m.) system of $\jpsi$, and ($\theta_{n},\phi_{n}$), ($\theta_{\pbar},\phi_{\pbar}$) are the polar and 
azimuthal angles of the neutron and anti-proton in their respective rest frames, defined in a right-handed system~\cite{Perotti2019}.
The parameters $\alpha_{\jpsi}$ and $\Delta \Phi$ describe the production of $\jpsi\rightarrow\Lambda\Lambdabar$, 
and $\alpha_{0}$ and $\alpha_{+}$ are the decay parameters of $\Lambda\to n\piz$ and $\Lambdabar\to \pbar\pip$, respectively.
%The components of these vectors are expressed using a right-handed coordinate system defined in Ref.~\cite{Perotti2019}, 
%$\alpha_{\jpsi}$ and $\Delta \Phi$ are the production parameters of $\jpsi\rightarrow\Lambda\Lambdabar$, and 
%$\alpha_{0}$ and $\alpha_{+}$ are the decay parameters of the $\Lambda$ and $\Lambdabar$ non-leptonic decay, respectively. 
The joint angular distribution of the decay is given by~\cite{Perotti2019},
\begin{equation}
\label{AngFunc}
\begin{split}
  \mathcal{W}(\xi;\omega) &= \mathcal{F}_{0}(\xi)+\alpha_{\jpsi} \mathcal{F}_{5}(\xi) + \alpha_{0}\alpha_{+}[\mathcal{F}_{1}(\xi) \\
                          &+\sqrt{1-\alpha_{\jpsi}^2}\cos(\Delta\Phi)\mathcal{F}_{2}(\xi)-\alpha_{\jpsi}\mathcal{F}_{6}(\xi)] \\
                          &+\sqrt{1-\alpha_{\jpsi}^2}\sin(\Delta \Phi)[-\alpha_{0}\mathcal{F}_{3}+\alpha_{+}\mathcal{F}_{4}],
\end{split}
\end{equation}
\noindent 
where the angular functions $\mathcal{F}_{i}(\xi)$ $(i=0, 1,..., 6)$ are described in Ref.~\cite{Faeldt2017}. 
In Eq.~\eqref{AngFunc}, the $\alpha_{0}\alpha_{+}$ terms arise from $\Lambda-\Lambdabar$ spin correlations,
while the $\alpha_{0}$ and $\alpha_{+}$ terms reflect the transverse polarization of $\Lambda$ and $\Lambdabar$, respectively.

%% BESIII
Details of the \mbox{BESIII} detector are given in Refs.~\cite{Ablikim2010, Huang2022}. 
The corresponding simulation, analysis framework, and software are presented in Refs.~\cite{RongGang2008, ZiYan2006, 5500be5449ac4841a9d6cfde58d0fef1}.
The Monte Carlo (MC) samples are generated with {\sc Geant}4-based~\cite{Agostinelli2003} software, which models the electron-positron collisions, particle decays, and detector response. 
A generic $\jpsi$ MC sample, with statistics matching the data, is used to study the potential background contribution. 
To eliminate experimental bias, The analysis is performed blindly with central values hidden~\cite{Klein2005} until the procedure is finalized.
%the analysis is performed blindly: central values are hidden using the method of Ref.~\cite{Klein2005} until event selection, fitting strategy and uncertainty evaluations are settled.
Signal MC samples are used to validate the analysis approaches and to study the systematic effects. 

Two oppositely charged tracks are detected in the main drift chamber (MDC).
The $\pbar$ and $\pip$ candidates are primarily selected as in Ref.~\cite{Ablikim2022a}.
A discrepancy between data and MC simulation is observed in the $\Lambdabar$ polar angle distribution, mainly due to the trigger efficiency in the endcap region. 
To minimize this effect, we require the absolute value of the cosine of the $\pbar$ polar angle must be less than 0.8.
%Therefore, the absolute value of the cosine of the $\pbar$ polar angle must be less than 0.8 to minimize the discrepancy between data and MC simulation.
%polar angle of $\pbar$ is required to fulfill $\lvert\cos\Theta_{\pbar}\rvert < 0.8$ to minimize the discrepancy between data and MC simulation.
%Due to the non-overlapping momentum coverage of anti-proton and pions in the kinematics, a charged track with momentum less than $0.5$~GeV/$c$ is assigned to be pion, otherwise anti-proton.
%Additionally, measurements of the specific ionization energy loss in the MDC and the flight time by the time-of-flight system are combined to perform the particle identification (PID),
%and an anti-proton candidate is identified by requiring the largest probability for the proton hypotheses.
%They are required to have the largest likelihood for the particle type selected among the pion, kaon, and proton hypotheses.
The $\Lambdabar$ candidate is reconstructed from the $\pbar$ and $\pip$ tracks via a vertex fit~\cite{Min2010},
and its invariant mass must lie within five times its resolution, where the mass resolution is determined in bins of the $\Lambdabar$ polar angle.
%The $\Lambdabar$ candidate must satisfy with $\lvert M_{\pbar\pip} - m_{\Lambda} \rvert < 5 \sigma_{1}$, where $M_{\pbar\pip}$ denotes the invariant mass of $\pbar\pip$ obtained from the vertex fit, 
%$\sigma_{1}$ is the corresponding mass resolution of $M_{\pbar\pip}$, which is a 8th polynominal function of polar angle of $\Lambdabar$ candidate and $m_{\Lambda}$ refers to the $\Lambda$ nominal mass~\cite{PDG}.
The decay length of the reconstructed $\Lambdabar$ is required to be positive.
To suppress backgrounds from $\jpsi\rightarrow\Sigma^{0}\bar{\Sigma}^{0}$, $\Sigma^{0}\bar{\Lambda}$, $\gamma\Lambda\Lambdabar$ and $\gamma\eta_{c}$, 
which include a $\Lambdabar$ in their final state, we furthermore require $\lvert M_{\Lambdabar}^{\rm rec} - m_{\Lambda} \rvert < 3\sigma_{\Lambdabar}^{\rm rec}$.
Here, $M_{\Lambdabar}^{\rm rec}=\sqrt{(E_{\rm cm}-E_{\Lambdabar})^{2}/c^{4}-P_{\Lambdabar}^{2}/c^{2}}$ is the recoiling mass of the $\Lambdabar$, obtained from the c.m. energy $E_{\rm cm}$, 
as well as the energy $E_{\Lambdabar}$ and momentum $P_{\Lambdabar}$ of the $\Lambdabar$ candidate from the vertex fit. 
The resolution $\sigma_{\Lambdabar}^{\rm rec}$ is determined in bins of the $\Lambdabar$ polar angle.

%$M_{\Lambdabar}^{\rm rec}=\sqrt{(E_{\rm cm}-E_{\Lambdabar})^{2}/c^{4}-P_{\Lambdabar}^{2}/c^{2}}$ is defined, where $E_{\rm cm}$ is the center-of-mass (c.m.) energy, $E_{\Lambdabar}$ and ${P_{\Lambdabar}}$ is the energy and momentum of $\Lambdabar$ candidate obtained from vertex fit, respectively.
%The recoil mass is required to be within $\lvert M_{\Lambdabar}^{\rm rec} - M_{\rm cent}^{\rm rec} \rvert < 3\sigma_{2}$, where $M_{\rm cent}^{\rm rec}$ is the center value of the recoil mass and $\sigma_{2}$ is the mass resolution of $M_{\Lambdabar}^{\rm rec}$ as the function of polar angle of $\pbar\pip$.
%To minimize the discrepancy between data and MC simulation, the polar angle $\theta_{\pbar}$ of reconstructed $\pbar$  is required to fulfill $\lvert\cos\theta_{\pbar}\rvert < 0.8$.

At least two photon candidates are required in the electromagnetic calorimeter (EMC).
Neutral showers in the EMC are primarily selected as in Ref.~\cite{Ablikim2024}.
%A photon candidate should have energy greater than $25$~MeV in the barrel region ($\lvert\cos\theta\rvert < 0.8$) or 
%$50$~MeV in the end-cap region ($0.86 < \lvert\cos\theta \rvert < 0.92$), where $\theta$ is the polar angle.
%The photon candidates must be separated apart the charged tracks with an opening angle greater than $20^\circ$ 
%for an anti-proton considering the annihilation effects, while greater than $10^\circ$ for other charged tracks. 
%To suppress the electronic noise and showers unrelated to the event, the photon candidates are required to have the EMC time difference from  the event start time within $[0, 700]$~ns.
To veto the showers from neutron (anti-neutron) interaction in the EMC, since the angle between neutron (anti-neutron) and $\Lambda (\Lambdabar)$  is very small (less than $7^{\circ}$),
the photon candidates should be separated from the $\Lambda$ ($\Lambdabar$) momentum direction 
by more than $10^\circ$ ($15^\circ$), where the $\Lambda$ ($\Lambdabar$) momentum direction is opposite to that of $\Lambdabar$ ($\Lambda$).
To further suppress the neutron-induced shower, a boosted decision tree (BDT) classifier~\cite{Therhaag2011} is constructed using shower shape variables including deposited energy, 
number of hits, second moment, Zernike moments, and deposition shape~\cite{Ablikim2024}.
Photons matched to truth information in signal MC are treated as signal, while others are treated as background.
A requirement on the BDT output $>0.15$ ($0.10$) is applied to ensure $90\%$ signal efficiency, with a corresponding $55\%$ ($45\%$) background rejection efficiency.
The $\piz$ candidates are reconstructed with a photon pair by performing a kinematic fit constraining their invariant mass to the $\piz$ nominal mass.
In case of multiple $\piz$ candidates from combinatorial effects, the one with the smallest kinematic fit $\chi^2$ is retained, requiring $\chi^2 < 100$.
%Due to combinatorial effects it is possible to have more than one unique $\pi^0$ candidate in a single event, only the candidate with the smallest $\chi^2$ is retained and a requirement $\chi^{2}<100$ is implemented.

\begin{figure}[hpt]
  \centering
  \begin{overpic}[scale=0.44]{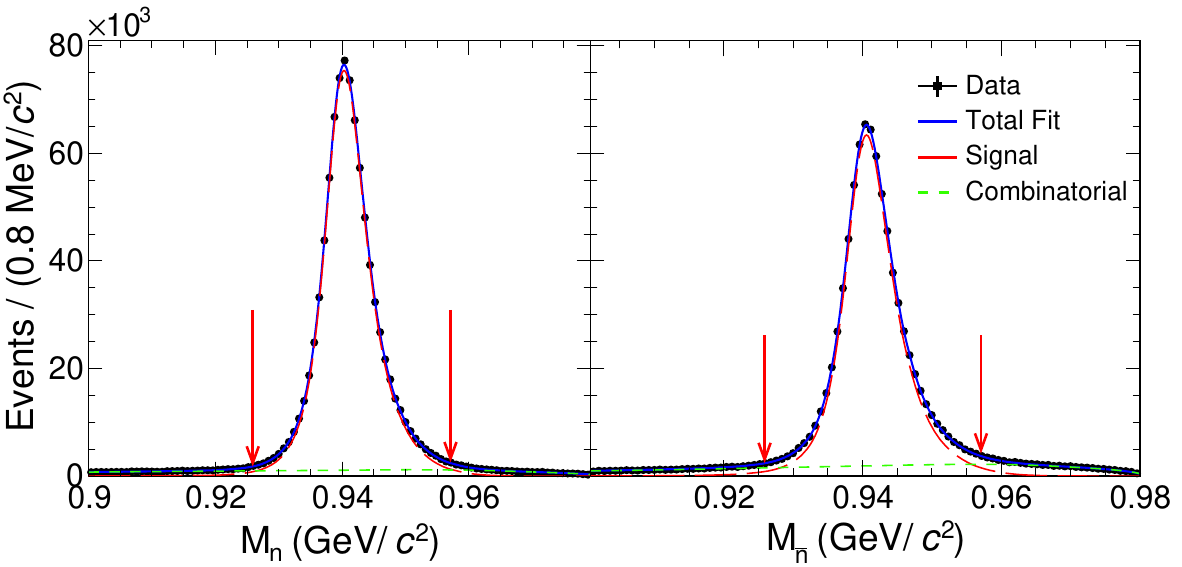}\end{overpic}
  %\begin{overpic}[scale=0.21]{Graph/FitMn_npi0.pdf}\put(75,58){(a)}\end{overpic}
  %\begin{overpic}[scale=0.21]{Graph/FitMn_nbarpi0.pdf}\put(74,58){(b)}\end{overpic}
  \caption{Distributions of the missing masses for $\Lambda\to n\piz$ and $\Lambdabar\to\nbar\piz$, respectively. 
%The dots with error bars are for data.
The blue solid curve represents the total fit result.
The red dashed and green dotted lines denote the signal shape and the combinatorial background contributions, respectively. 
The red arrows indicate the signal region.}
 \label{fig:hyperondecay}
\end{figure}

Because of the limited detection efficiency for neutrons and antineutrons, they are not reconstructed directly. 
Instead, their four-momenta are inferred as $\bm{P}_{n}=\bm{P}_{\Lambda}-\bm{P}_{\piz}$, 
where $\bm{P}_{n}$, $\bm{P}_{\Lambda}$ and $\bm{P}_{\piz}$ are the four-momenta of neutron, $\Lambda$ and $\piz$ candidate, respectively.
To improve the resolution, the momentum and energy of $\Lambda$ are set to $\vec{p}_{\Lambda}=-\hat{p}_{\Lambdabar}\sqrt{E^{2}_{\rm cm}/c^{2}/4-m_{\Lambda}^{2}c^{2}}$ and $E_{\Lambda}=E_{\rm cm}/2$, respectively, 
%Since all the other final state particles are detected, it allows for the reconstruction of the neutron's kinematics from calculation $\bm{P}_{n}=\bm{P}_{\Lambda}-\bm{P}_{\piz}$, where $\bm{P}$ indicates the four momentum.
%With reconstructing the $\Lambdabar$, to improve the signal resolution, the momentum and energy of $\Lambda$ are denoted by $\vec{p}_{\Lambda}=-\hat{p}_{\Lambdabar}\sqrt{E^{2}_{\rm beam}/c^{2}-m_{\Lambda}^{2}c^{2}}$ and $E_{\Lambda}=\sqrt{|\vec{p}_{\Lambda}|^{2}c^{2}+m^{2}_{\Lambda}c^{4}}$, respectively,
where $\hat{p}_{\Lambdabar}$ is the momentum direction of the $\Lambdabar$ in the $e^{+}e^{-}$ rest frame.
The neutron is identified by the missing mass calculated from $\bm{P}_{n}$, which shows a clear peak near the nominal neutron mass with low background (Fig.~\ref{fig:hyperondecay}).
%The neutron signal is identified with its missing mass extracted from $\bm{P}_{n}$, as shown in Fig.~\ref{fig:hyperondecay}, 
%where a prominent signal peak with low background is observed near the neutron mass.
%Therefore, the signal is identified by the missing mass, as shown in Fig.~\ref{fig:hyperondecay} with a prominent signal peak in the neutron vicinity and a low level background.
Studies of MC samples and data in the neutron sideband show that background in the signal region mainly arises from signal events with misreconstructed $\piz$ candidates, i.e., combinatorial background.
%By performing detailed studies on the MC simulation samples and data in the neutron mass sideband region,
%the background candidates in the signal region are found to be the signal events with misreconstructed $\piz$ candidates, namely combinatorial background.  
The combinatorial background does not peak in the neutron invariant mass distribution.
%It is referred to as combinatorial background hereafter.
The other backgrounds, stemming from processes with one or two extra photons in the final state~\cite{Zhou2021}, contribute less than $0.5\%$ and are neglected in the nominal analysis. 

%To extract the signal yields, unbinned maximum likelihood fits are carried out on the distributions of neutron missing mass, as shown in  Fig.~\ref{fig:hyperondecay}.
%In the fit,
%%Signal yields are obtained from an unbinned maximum likelihood fit on the missing mass.
%%In the fit shown in Fig.~\ref{fig:hyperondecay}, 
%the signal is described with a MC-simulated shape convoluted with a Gaussian function to account for the resolution difference between data and MC simulation, 
%and the combinatorial background is described by the shape from the corresponding MC sample. %described by the MC-simulated shape.
%The fit yields  $872449\pm982$ signal events and a purity of $95.3\%$ in the signal range  $[0.926, 0.957]$~GeV/$c^2$ for the decay $\Lambda\to n \piz$, 
%while $759770\pm963$ signal events and a purity of $91.1\%$ for the decay $\Lambdabar\to \nbar \piz$.

To extract the production and decay parameters, a simultaneous fit on the joint angular distribution is performed on the selected events in the neutron/anti-neutron missing-mass signal range $[0.926, 0.957]$~GeV/$c^2$, 
with the production parameters $\alpha_{\jpsi}$ and $\Delta\Phi$ shared between the two charge conjugated channels.
In the fit, the probability density function for the $i$-th event with helicity angle $\xi_i$ is parameterized with the four parameters $\omega$, 
%is written as, %of the four unknown parameters $\omega$ can be defined in terms of the helicity angles $\xi$
\begin{equation}
	\mathcal{P}(\xi_i;\omega) = \mathcal{W}(\xi_i;\omega)\varepsilon(\xi_i)/\mathcal{N}(\omega),
\end{equation}
\noindent
where the normalization factor $\mathcal{N}(\omega)$ is obtained by $\mathcal{N}(\omega) = \frac{1}{M} \sum_{j = 1}^{M} \frac{\mathcal{W}(\xi_j;\omega)}{\mathcal{W}(\xi_j;\omega_{\rm gen})}$ 
with a signal MC sample generated with known parameters $\omega_{\rm gen}$~\cite{Ablikim2022b,Ablikim2024} and size $M$, similar to that in Ref.~\cite{Ablikim2024}. 
The sum runs over all $M$ signal MC events, corresponding to 50 times the data yield.
The log-likelihood function for $N$ selected candidates is
\begin{equation}
	\small
	\mathcal{S} = -\sum_{i = 1}^N\ln \mathcal{P}(\xi_i;\omega) + \frac{N_{\rm bkg}}{N_{\rm miss}} \sum_{i}^{N_{\rm miss}} \ln \mathcal{P} (\xi_i ; \omega),
\end{equation}
\noindent
where the second term accounts for the contribution from the combinatorial background, estimated using $N_{\rm miss}$ MC-simulated background events.
To determine the background yield $N_{\rm bkg}$, unbinned maximum likelihood fits are carried out on the distributions of the neutron missing mass, as shown in  Fig.~\ref{fig:hyperondecay}.
In the fit, the signal is modeled with an MC-simulated shape convolved with a Gaussian function to account for the resolution difference between data and MC simulation,
while the combinatorial background is modeled by the shape from the corresponding MC sample.
%The fit yields background events $N_{\rm bkg}=42925\pm293$ and a background level of $4.7\%$ in the signal range for the decay $\Lambda\to n \piz$,
%while $N_{\rm bkg}=74567\pm379$ and a background level of $8.9\%$ for the decay $\Lambdabar\to \nbar \piz$.
The fit yields $872450\pm982$ signal events with a purity of $95.3\%$ for the decay $\Lambda\to n \piz$,
and $759770\pm963$ events with a purity of $91.1\%$ for the decay $\Lambdabar\to \nbar \piz$.
%still left in the final event sample. 
%Since small differences between data and MC lead to substantial variations in $\alpha_{\jpsi}$, it is fixed to the value in Ref.~\cite{Ablikim2022b}.

\begin{figure}[hpt]
  \centering
  \begin{overpic}[scale=0.245]{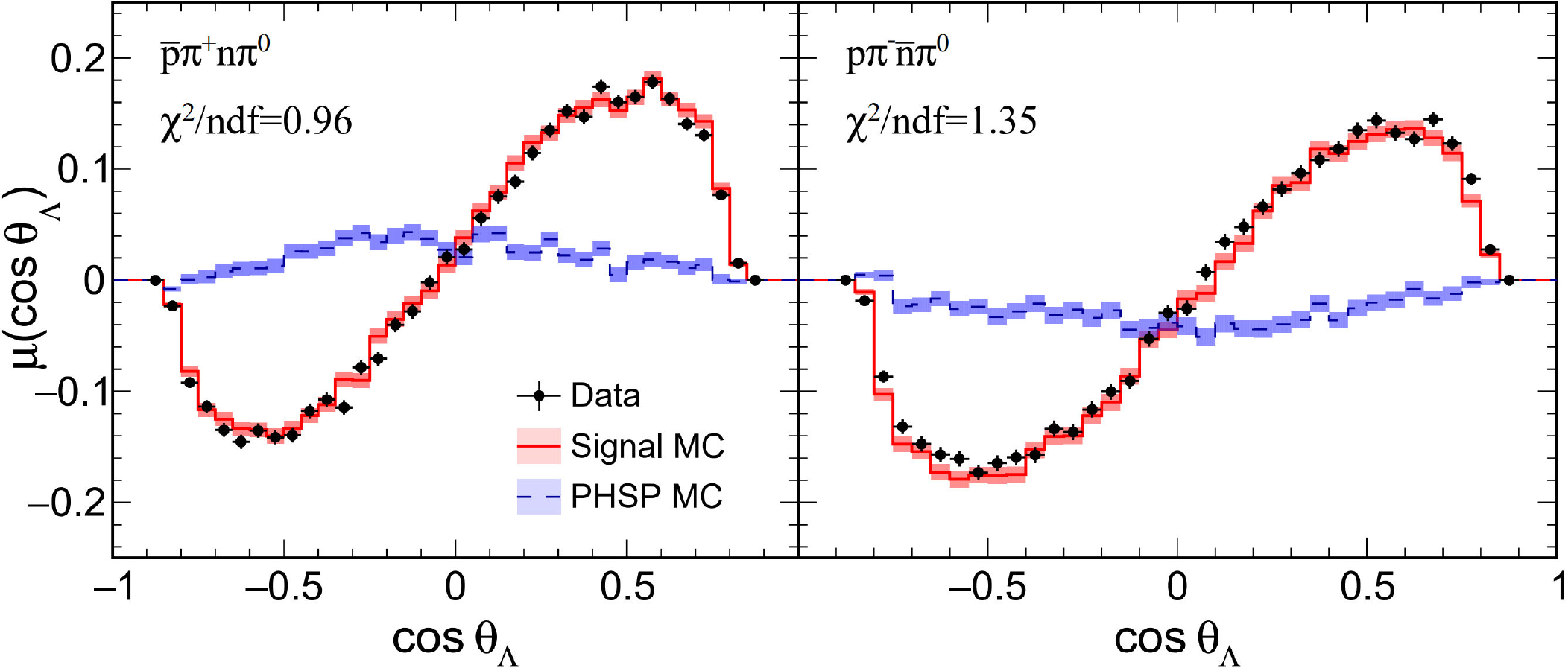}\end{overpic}
  %\begin{overpic}[scale=0.132]{Graph/Polarization.pdf}\end{overpic}
  \caption{Distributions of $\mu(\cos\theta_{\Lambda})$ as a function of $\cos\theta_{\Lambda}$.
The dots with error bars are data, the red solid lines are the signal MC sample and the blue dashed line represent the distributions without polarization from the PHSP MC sample.
The red and blue bands indicate the statistical uncertainties of signal MC and PHSP MC samples, respectively.}
 \label{fig:polarization}
\end{figure}

The parameters $\omega$ are determined by a maximum likelihood fit, minimizing $\mathcal{S}$ with the TMinuit~\cite{Hatlo2005} package.
The parameter $\alpha_{\jpsi}$ is fixed to the value reported in Ref.~\cite{Ablikim2022b}, a consequence of the data-MC discrepancy mentioned earlier.
The fit results are summarized in Table~\ref{tab:fitresults}.
The correlation coefficients are $\rho(\alpha_{0},\bar{\alpha}_{0})=-0.005$ and $\rho(\alpha_{-}, \alpha_{+})=-0.026$.
The moment $\mu$, defined as
\begin{equation}
  \mu(\cos\theta_{\Lambda})=\frac{m}{N}\sum_{i=1}^{N_{k}}(\sin\theta_{n}^{i}\sin\phi_{n}^{i}+\sin\theta_{\pbar}^{i}\sin\phi_{\pbar}^{i}),
\end{equation}
\noindent
is proportional to the product of the $\Lambda$ polarization and its decay asymmetry. The quantity is evaluated across $m=40$ bins in $\cos\theta_{\Lambda}$. 
For each bin $k$, containing $N_k$ events, the index $i$ denotes the $i$-th event within it.
%Comparing the data to the phase space (PHSP) MC sample, as shown in Fig.~\ref{fig:polarization}, there is a significant polarization of the $\Lambda$ hyperons produced in $\jpsi\to\Lambda\Lambdabar$ decays.
Comparison of the data with the phase-space (PHSP) MC sample reveals a significant polarization of the $\Lambda$ hyperons in $\jpsi\to\Lambda\Lambdabar$ decays, as shown in Fig.~\ref{fig:polarization}.

\begin{table}[hbtp]
	\centering
	\scriptsize
	\caption[]{Summaries of the production and decay parameters, the $CP$ asymmetry $A_{CP}^{-}=(\alpha_{-}+\alpha_{+})/(\alpha_{-}-\alpha_{+})$ 
	and $A_{CP}^{0}=(\alpha_{0}+\bar{\alpha}_{0})/(\alpha_{0}-\bar{\alpha}_{0})$, the average decay parameters $\langle\alpha_{0}\rangle=(\alpha_{0}-\bar{\alpha}_{0})/2$ 
	and $\langle\alpha_{-}\rangle=(\alpha_{-}-\alpha_{+})/2$, as well as the ratios $\az/\am$ and $\azbar/\alp$. 
    The first uncertainties are statistical and the second ones systematic.}
	\renewcommand\arraystretch{1.5}
	\label{tab:fitresults}
 \resizebox{\linewidth}{!}{
	\begin{tabular}{lrr}
		\hline \hline
		Parameter & This work & Previous result   \\
		\hline
$\Delta\Phi$ (rad) & $ 0.748\pm0.006\pm 0.004$ & $0.7521 \pm 0.0042 \pm 0.0066$~\cite{Ablikim2022b}   \\
$\alpha_{-}$ & $ 0.756 \pm 0.008 \pm 0.003$ & $0.7519\pm0.0036\pm0.0024$~\cite{Ablikim2022b}   \\
$\alpha_{+}$ & $ -0.764 \pm 0.008 \pm 0.001$ & $-0.7559\pm0.0036\pm0.0030$~\cite{Ablikim2022b}   \\
$\alpha_{0}$ & $ 0.668 \pm 0.007 \pm 0.002$ & $0.670 \pm 0.009 ^{+0.009}_{-0.008}$~\cite{Ablikim2024}   \\
$\bar{\alpha}_{0}$ & $ -0.677 \pm 0.007 \pm 0.003$ & $-0.668 \pm 0.008 ^{+0.006}_{-0.008}$~\cite{Ablikim2024}   \\
$A_{CP}^{-}$ & $ -0.005 \pm 0.007 \pm 0.002$ & $-0.0025\pm0.0046\pm0.0012$~\cite{Ablikim2022b}    \\
$A_{CP}^{0}$ & $ -0.006 \pm 0.007 \pm 0.002$ & $0.001 \pm 0.009 ^{+0.005}_{-0.007}$~\cite{Ablikim2024}    \\
$\langle\alpha_{0}\rangle$ & $ 0.672 \pm 0.002 \pm 0.002$ & $0.673 \pm 0.006 ^{+0.007}_{-0.006}$~\cite{Ablikim2024}    \\
$\langle\alpha_{-}\rangle$ & $ 0.760 \pm 0.002 \pm 0.002$ & $0.7542 \pm 0.0010 \pm 0.0024$~\cite{Ablikim2022b}    \\
$\alpha_{0}/\alpha_{-}$ & $ 0.884 \pm 0.013 \pm 0.006$ & $0.877 \pm 0.015 ^{+0.014}_{-0.010}$~\cite{Ablikim2024}    \\
$\bar{\alpha}_{0}/\alpha_{+}$ & $ 0.885 \pm 0.013 \pm 0.004$ & $0.863 \pm 0.014 ^{+0.012}_{-0.008}$~\cite{Ablikim2024}    \\
		\hline 
        \hline
	\end{tabular}
 }
\end{table}

The sources of systematic uncertainties are separated into four categories: event reconstruction, background estimation, fit procedure and residual Data/MC discrepancy.
For event reconstruction, the effects from $p$, $\pbar$, and $\pi^\pm$ tracking, PID, and $\Lambda/\Lambdabar$ reconstruction are evaluated with a control sample of $\jpsi \to \Lambda (\to p\pi^-) \bar{\Lambda} (\to \pbar \pi^+)$.
The $\piz$ reconstruction efficiency is studied with $\jpsi \to \Sigma^+ (\to p\piz)\bar{\Sigma}^- (\to \pbar\piz)$, which has a final-state topology similar to the signal.
Selection-related effects, such as the polar angle requirement of $\pbar$, the missing-mass window, and the $\chi^2$ of the $\piz$ kinematic fit, are estimated by varying their values around the nominal ones and repeating the fits.
%The systematic uncertainties associated with $p$, $\pbar$ and $\pi^\pm$ tracking, PID and $\Lambda/\Lambdabar$ reconstruction are investigated 
%with a control sample $\jpsi \to \Lambda (\to p\pi^-) \bar{\Lambda} (\to \bar{p} \pi^+)$.
%The uncertainty of $\piz$ reconstruction is investigated by studying the control sample $\jpsi \to \Sigma^+ (\to p\piz) \bar{\Sigma}^{-} (\to \pbar\piz)$, 
%which has a similar final state topology than the signal.
%The systematic uncertainties related to the selection criteria including the requirements of the decay length, invariant masses and recoil mass of $\Lambdabar$, the polar angle of $\pbar$, 
%The systematic uncertainties related to the selection criteria including the requirements of the polar angle of $\pbar$, 
%the missing mass and the $\chi^2$ of the $\piz$ kinematic fit are studied by varying their required values around the nominal ones and repeating the fits.
Background-related uncertainties are evaluated in two ways: a $\pm 1\sigma$ variation of the combinatorial background yield in the fit, and the inclusion of the dominant neglected background term, as determined from MC simulation, in an alternative fit.
%The uncertainty associated with the combinatorial background is obtained by varying its yield by $\pm 1\sigma$ in the fit. % which is determined from the fit to the missing mass distribution.
%The uncertainty associated with the neglected background contributions in the fit is studied by performing an alternative fit including the dominant background term in the log-likelihood function, 
%where this background term is evaluated with the corresponding background MC samples.  
%which is determined by the difference between nominal result and the fit result with adding a new term $\sum_{i}^{N_{\rm rb}} \ln \mathcal{P}(\xi_i;\omega)$ to the likelihood function, where the $N_{\rm rb}$ is the number of resonant backgrounds from inclusive MC sample.
The uncertainty of fit procedure is tested with 50 sets of toy MC samples generated using the best-fit parameters.
%The uncertainty associated with the fit procedure is studied by performing individual fits on 50 sets of toy MC samples generated with the parameters obtained in this analysis. 
Since the input and output are consistent within statistical fluctuations, no additional uncertainty is assigned to the fit procedure.
%Then the sum of the results offset and the uncertainties of mean value of pull distribution is taken as the uncertainties.
%To estimate the systematic uncertainty of the fit procedure, 50 sets of toy MC samples are generated. Each set is fitted to obtain the distribution of the output parameters. 
%The sum of the mean values of the pull distributions and their uncertainties are regarded as systematic uncertainties.
The fixed input of $\alpha_{\jpsi}$ is studied by repeating the fit with $\alpha_{\jpsi}$ randomly varied within one standard deviation for 100 iterations. The standard deviation of the output is assigned as the corresponding uncertainty. 
Finally, the residual data/MC discrepancy is evaluated by reweighting the MC to match the $\Lambdabar$ polar-angle distribution in data; the shift in the fitted result is taken as the associated uncertainty.
%The systematic uncertainty associated with the data/MC discrepancy is evaluated by reweighting the MC to match the data in the distribution of the polar angle of $\Lambdabar$. 
%The resulting weights are applied in the fit, and the difference between this weighted fit result and the nominal result is assigned as the systematic uncertainty.
%is studied by randomly changing the $\alpha_{\jpsi}$ value within one standard deviation 100 times in the fit.
%Then a Gaussian function is used to fit the 100 trials and the width of the Gaussian function is estimate to the systematic uncertainties.
The absolute contributions from various sources are summarized in Table~\ref{tab:SysUncSource}.
The total systematic uncertainty for each parameter is obtained by summing these individual contributions in quadrature.

\begin{table}[hbtp]
	\centering
	\scriptsize
	\caption[]{Absolute systematic uncertainties of the measured parameters (in units of $10^{-3}$).
	%the parameter $\Delta\Phi$ (rad),  
	%the decay parameters $\alpha_{-}$, $\alpha_{+}$, $\alpha_{0}$, $\bar{\alpha}_{0}$, 
	%the $CP$ asymmetry $A_{CP}^{-}=(\alpha_{-}+\alpha_{+})/(\alpha_{-}-\alpha_{+})$, $A_{CP}^{0}=(\alpha_{0}+\bar{\alpha}_{0})/(\alpha_{0}-\bar{\alpha}_{0})$, 
	%the average decay parameters $<\alpha_{0}>$ and $<\alpha_{-}>$,
	%and the ratios of $\az/\am$ and $\azbar/\alp$. 
	%The systematic uncertainties of the parameters in the last six rows of the table are calculated using the same method as that of the decay parameters.
	%All numerical values are expressed in units of $10^{-3}$. 
     The ``Event Reco'', ``Bkg Est'', ``Fit Meth'' and ``Data/MC Diff.'' represent the systematic uncertainty associated with event reconstruction, background estimation, fit method, and residual data/MC discrepancy, respectively.
	}
	\renewcommand\arraystretch{1.5}
	\label{tab:SysUncSource}
 	\resizebox{\linewidth}{!}{
	%\begin{tabular}{lccccccccc}
	\begin{tabular}{ccccc|c}
	\hline 
	\hline
      Parameter & \makecell{Event\\ Reco} & \makecell{Bkg\\ Est} & \makecell{Fit\\ Meth} & \makecell{Data/MC\\ Diff.} & Total \\
      \hline
      $\Delta\Phi$ (rad) & 1.3 & 2.7 & 2.3 & 2.1 & 4.3 \\
      $\alpha_{-}$ & 1.9 & 1.8 & 0.3 & 0.9 & 2.8 \\
      $\alpha_{+}$ & 0.9 & 0.7 & 0.3 & 0.3 & 1.2 \\
      $\alpha_{0}$ & 1.6 & 0.4 & 0.6 & 0.6 & 1.9 \\
      $\bar{\alpha}_{0}$ & 2.4 & 0.4 & 0.6 & 2.0 & 3.2 \\
      $A_{CP}^{-}$ & 1.5 & 0.8 & 0.0 & 0.8 & 1.9 \\
      $A_{CP}^{0}$ & 2.0 & 0.1 & 0.1 & 1.0 & 2.2 \\
      $\langle\alpha_{0}\rangle$ & 1.1 & 0.4 & 0.6 & 1.3 & 1.9 \\
      $\langle\alpha_{-}\rangle$ & 0.9 & 1.2 & 0.3 & 0.3 & 1.6 \\
      $\alpha_{0}/\alpha_{-}$ & 5.0 & 2.7 & 0.5 & 1.8 & 5.9 \\
      $\bar{\alpha}_{0}/\alpha_{+}$ & 3.4 & 1.2 & 0.4 & 2.2 & 4.3 \\

      %Parameter & \makecell{Event\\ reconstruction} & \makecell{Background\\ estimation} & \makecell{Fit\\ method} & Total \\
      %\hline
      %$\Delta\Phi$ (rad) & 1.3 & 2.7 & 3.1 & 4.3 \\
      %$\alpha_{-}$ & 1.9 & 1.8 & 0.9 & 2.8 \\
      %$\alpha_{+}$ & 0.9 & 0.7 & 0.5 & 1.2 \\
      %$\alpha_{0}$ & 1.6 & 0.4 & 0.9 & 1.9 \\
      %$\bar{\alpha}_{0}$ & 2.4 & 0.4 & 2.1 & 3.2 \\
      %$A_{CP}^{-}$ & 1.5 & 0.8 & 0.8 & 1.9 \\
      %$A_{CP}^{0}$ & 2.0 & 0.1 & 1.0 & 2.2 \\
      %$\langle\alpha_{0}\rangle$ & 1.1 & 0.4 & 1.4 & 1.9 \\
      %$\langle\alpha_{-}\rangle$ & 0.9 & 1.2 & 0.4 & 1.6 \\
      %$\alpha_{0}/\alpha_{-}$ & 5.0 & 2.7 & 1.9 & 5.9 \\
      %$\bar{\alpha}_{0}/\alpha_{+}$ & 3.4 & 1.2 & 2.3 & 4.3 \\

	\hline 
    \hline
	\end{tabular}
 }
\end{table}

\begin{figure}[hpt]
  \centering
  %\begin{overpic}[scale=0.4]{Graph/FinalResult2D.pdf}\end{overpic}
  \begin{overpic}[scale=0.20]{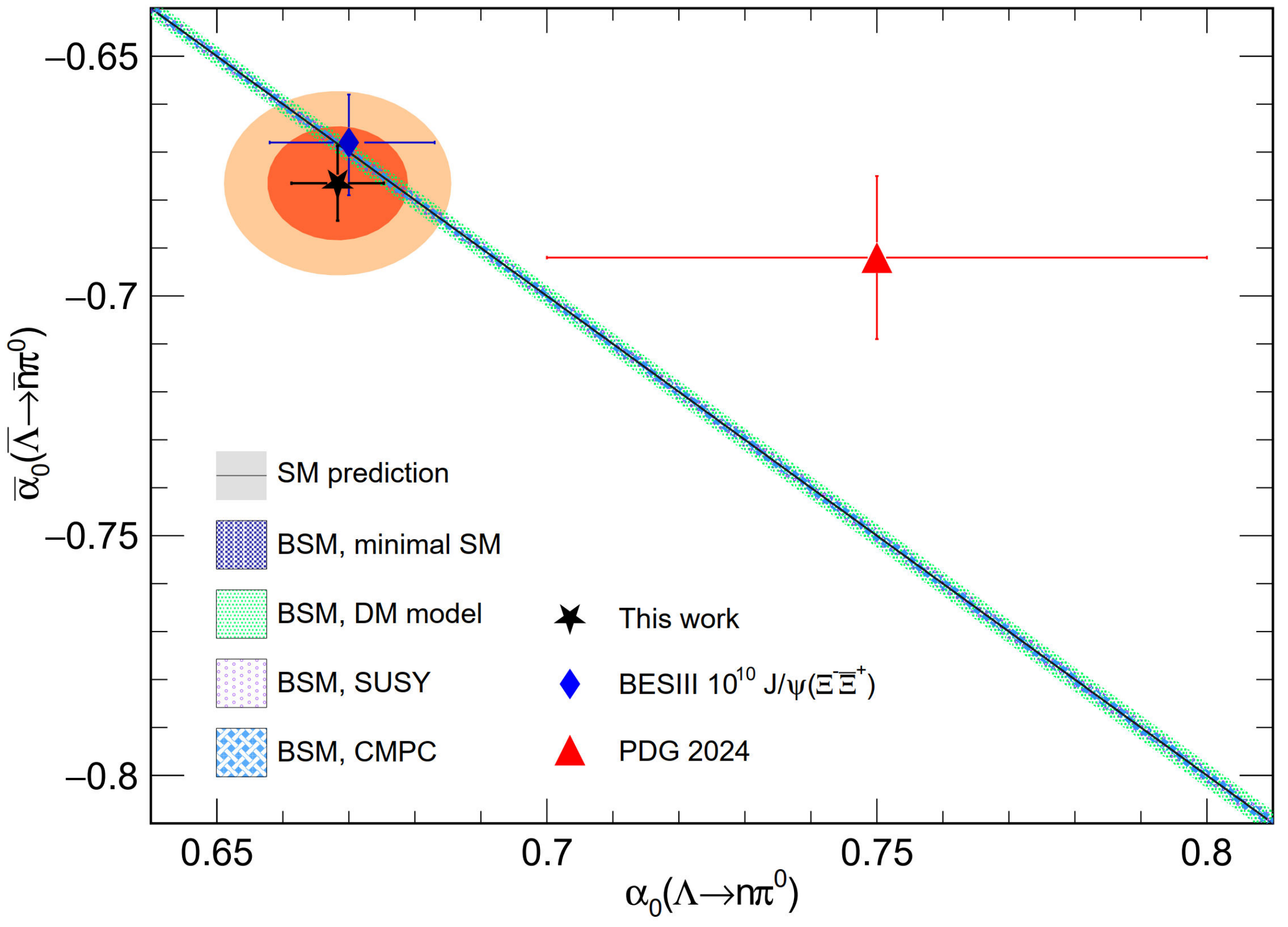}\end{overpic}
  \caption{Two dimensional distribution of the decay parameters $\bar{\alpha}_{0}$ versus $\alpha_{0}$. 
  The black star with error bars denotes the results measured in this work and the orange contours correspond to the $68\%/95\%$ confidence levels.
  The blue diamond shows the results of $\alpha_{0}$ and $\bar{\alpha}_{0}$ by the process $J/\psi\to\Xi^{-}\bar{\Xi}^{+}$ with 10 billion $J/\psi$ events~\cite{Ablikim2024}.
  %The blue quadrilateral star indicates the result for $\bar{\alpha}_{0}$ by process $J/\psi\to\Lambda\Lambdabar$ with 1.3 billion $J/\psi$ events~\cite{BESIII:2018cnd}.
  The red triangle represents the PDG 2024 values~\cite{PDG}.
  The black line with gray band represents the prediction from the SM~\cite{Tandean2003, HE20221840}.
  The blue, green, violet and azure bands indicate the results predicted by minimal standard model~\cite{PhysRevD.52.5257}, 
  dark matter model~\cite{he2025btoksfinvisibledarkmatter}, supersymmetry~\cite{PhysRevD.61.071701} and chromomagnetic-penguin contributions~\cite{HE20221840}, respectively.
  %Results of the $\Lambda$ neutral decay parameters $\alpha_{0}$ and $\bar{\alpha}_{0}$ from different measurements. 
  %The black dot represents the PDG 2024 value of $\alpha_{0}$~\cite{PDG}, the pink dot represents the result for $\bar{\alpha}_{0}$ by process $J/\psi\to\Lambda\Lambdabar$ 
  %with 1.3 billion $J/\psi$ events~\cite{BESIII:2018cnd}, the blue dots represents the results of both $\alpha_{0}$ and $\bar{\alpha}_{0}$ 
  %by process $J/\psi\to\Xi^{-}\bar{\Xi}^{+}$ with 10 billion $J/\psi$ events~\cite{Ablikim2024}, and the red dots indicates the results from this analysis. 
  }
 \label{fig:FinalResult}
\end{figure}

Based on $(10087\pm44)\times 10^{6}$ $\jpsi$ events collected with the BESIII detector, 
we perform a full angular analysis of the decay $\jpsi \to \Lambda (\to n \piz) \Lambdabar (\to \pbar \pip)$ and its charge conjugated mode.
%a full angular analysis of the decay $\jpsi \to \Lambda (\to n \piz) \Lambdabar (\to \pbar \pip)$ and its charge conjugated mode has been performed.
%As shown in Table~\ref{tab:fitresults}, 
The neutral decay parameters $\alpha_0$ and $\bar{\alpha}_0$ are  %to be $\alpha_0= 0.668 \pm 0.007 \pm 0.002$, $\bar{\alpha}_0 =-0.677 \pm 0.007 \pm 0.005$, 
consistent with previous measurements~\cite{Ablikim2024, BESIII:2018cnd}, with precisions improved by 43\% and 27\%, respectively, as shown in Fig.~\ref{fig:FinalResult}.
This improvement is primarily reflected in the reduced systematic uncertainties.
%, owing to the simplified selection criteria and more refined MC efficiency corrections employed in this analysis.
We also determine $\Delta\Phi$ and the charged decay parameters $\alpha_{-}$ and $\alpha_{+}$, establishing nonzero polarization in $\jpsi\to\Lambda\Lambdabar$, 
%The analysis has also yielded the production parameter $\Delta\Phi$ and the decay parameters of the charged $\Lambda$ decay modes, $\alpha_{-}$ and $\alpha_{+}$, 
%$\Delta\Phi = 0.748 \pm 0.006 \pm 0.004$ rad, $\alpha_{-} = 0.756 \pm 0.008 \pm 0.007$ and $\alpha_{+} = -0.764 \pm 0.008 \pm 0.002$, individually, 
which are all consistent with the previous results~\cite{Ablikim2022b, BESIII:2018cnd, Ablikim2024, PhysRevD.108.L031106}.
The derived $CP$ asymmetries $A_{CP}^{-}$ and $A_{CP}^{0}$ are consistent with zero, with $A_{CP}^{0}$ obtained at the highest precision to date (29\% improvement).
The ratios of the $\Lambda$ decay parameters between the isospin conjugated decay modes $\az/\am$ and $\azbar/\alp$, deviate from unity by $8.1\sigma$ and $8.5\sigma$, respectively, confirming the $\Delta I = 3/2$ contributions.
% have been obtained, %to be $\az/\am= 0.884\pm0.013\pm0.007$ and $\azbar/\alp=0.885\pm0.013\pm0.004$, respectively, 
%which are consistent with the previous results but with improved precision~\cite{Ablikim2024}.
%The obtained ratios $\az/\am$ and $\azbar/\alp$ deviate from unity by $8.1\sigma$ and $8.5\sigma$, respectively, 
%confirming that the $\Delta I = 3/2$ transition is involved in both the decays of $\Lambda$ and $\Lambdabar$.
Combinning the averages of $\langle\alpha_{0}\rangle$, $\langle\alpha_{-}\rangle$, the decay rates 
%Using the averages of the $\langle\alpha_{0}\rangle$ and $\langle\alpha_{-}\rangle$  with combinations of the decay rates 
%$<\alpha_{0}>=0.672\pm0.005\pm0.003$ and $<\alpha_{-}>=0.760\pm0.006\pm0.002$ with combinations of the decay rates 
$\Gamma(\Lambda\to p\pim)$, $\Gamma(\Lambda\to n\piz)$~\cite{PDG} and the $N-\pi$ scattering phase shift~\cite{Meissner:2017jjx}, 
the ratio of $\Delta I=3/2$ to $\Delta I=1/2$ transitions in the $S$ wave is determined to be $S_{3}/S_{1}=0.033\pm0.002\pm0.001\pm0.004$, 
and in the $P$ wave to be $P_{3}/P_{1}=-0.067\pm0.008\pm0.002\pm0.004$, in line with Ref.\cite{Salone2022}. Here, the first uncertainties are statistical, 
the second systematic, and the third arise from the input parameters~\cite{PDG, Meissner:2017jjx}. 
The $S$-wave ratio agrees with $\mathrm{Re}(A_{2})/\mathrm{Re}(A_{0})$ in $K\to\pi\pi$ within the uncertainty, while the $P$-wave ratio shows a discrepancy.
Building upon previous work~\cite{Ablikim2024}, this analysis achieves the most precise measurement of neutral decays of $\Lambda$ to date.
It also demonstrates the necessity of including $\Delta I=3/2$ amplitudes, and provide important benchmarks for future high-precision $e^+e^-$ studies.
%demonstrating particular improvements in the systematic uncertainty compared to previous analyses~\cite{Ablikim2024},
%which provides significant prospects for future high-precision electron-positron collider experiments.
%Moreover, the study demonstrates that neglecting the $\Delta I = 3/2$ contribution in $\Lambda$ decays is unjustified, will be essential for refining theoretical approaches and for guiding future high-precision studies.

%This analysis provides valuable results relevant to the study of the underlying dynamics of hyperon non-leptonic decays.
%In addition, we obtain the value of $CP$ violation for both $\Lambda$ charged and neutral decays $A_{CP}^{-} = -0.005 \pm 0.007 \pm 0.006$ and $A_{CP}^{0} = -0.006 \pm 0.007 \pm 0.004$,
%which are compatible with zero, thereby, indicating a non-$CP$-violation scenario.
%The next generation of charm factories~\cite{Achasov2023, Bondar2013} will greatly improve the accuracy of the $CP$-violating measurements, and shed light on the mechanism of $CP$ violation in the baryon sector.

\textbf{Acknowledgement}

The BESIII Collaboration thanks the staff of BEPCII (https://cstr.cn/31109.02.BEPC), the IHEP computing center and 
the supercomputing center of the University of Science and Technology of China (USTC) for their strong support.
This work is supported in part by National Key R\&D Program of China under Contracts Nos. 2023YFA1606000, 2023YFA1606704, 2023YFA1609400; 
National Natural Science Foundation of China (NSFC) under Contracts Nos. 
11635010, 11935015, 11935016, 11935018, 12025502, 12035009, 12035013, 12061131003, 12192260, 12192261, 12192262, 12192263, 12192264, 12192265, 12221005, 12225509, 12235017, 12361141819, 12122509, 12105276; 
the Chinese Academy of Sciences (CAS) Large-Scale Scientific Facility Program; 
the Strategic Priority Research Program of Chinese Academy of Sciences under Contract No. XDA0480600; 
CAS under Contract No. YSBR-101; 100 Talents Program of CAS; The Institute of Nuclear and Particle Physics (INPAC) and Shanghai Key Laboratory for Particle Physics and Cosmology; 
ERC under Contract No. 758462; German Research Foundation DFG under Contract No. FOR5327; Istituto Nazionale di Fisica Nucleare, Italy; 
Knut and Alice Wallenberg Foundation under Contracts Nos. 2021.0174, 2021.0299; Ministry of Development of Turkey under Contract No. DPT2006K-120470; 
National Research Foundation of Korea under Contract No. NRF-2022R1A2C1092335; National Science and Technology fund of Mongolia; 
Polish National Science Centre under Contract No. 2024/53/B/ST2/00975; STFC (United Kingdom); 
Swedish Research Council under Contract No. 2019.04595; U. S. Department of Energy under Contract No. DE-FG02-05ER41374.
This paper is also supported by the Joint Large-Scale Scientific Facility Funds of the NSFC and CAS under Contracts Nos. U2032111;
Beijing Natural Science Foundation of China (BNSF) under Contract No. IS23014.

\bibliographystyle{apsrev4-2.bst}
\bibliography{main}

\end{document}